\documentstyle[12pt,epsf]{article}
\newcommand{\la}[1]{\label{#1}}

\newcommand{\be}{\begin{equation}}
\newcommand{\ee}{\end{equation}}
\newcommand{\ba}{\begin{eqnarray}}
\newcommand{\ea}{\end{eqnarray}}
\newcommand{\rmi}[1]{{\mbox{\scriptsize #1}}}
 
\newcommand{\nr}[1]{(\ref{#1})}

\newcommand{\bfs}{\mbox{\bf s}}

\newcommand{\bfb}{\mbox{\bf b}}

\newcommand{\roots}{\sqrt{s}}

\newcommand{\psat}{p_{\rmi{sat}}}

\newcommand{\fr}[2]{{\frac{#1}{#2}}}

\def\lsim{\raise0.3ex\hbox{$<$\kern-0.75em\raise-1.1ex\hbox{$\sim$}}}
\def\gsim{\raise0.3ex\hbox{$>$\kern-0.75em\raise-1.1ex\hbox{$\sim$}}}

\begin{document}
  
\begin{titlepage}
\begin{flushright}
20 September 2000\\
JYFL-3/00\\
HIP-2000-45/TH\\
hep-ph/0009246\\
\end{flushright}
\begin{centering}
\vfill

{\bf 

          CENTRALITY DEPENDENCE OF MULTIPLICITIES IN ULTRARELATIVISTIC
                     NUCLEAR COLLISIONS 
}

\vspace{0.5cm}
 K.J. Eskola,$^{\rm a,c,}$\footnote{kari.eskola@phys.jyu.fi}
 K. Kajantie$^{\rm b,}$\footnote{keijo.kajantie@helsinki.fi}  and
K. Tuominen$^{\rm a,}$\footnote{kimmo.tuominen@phys.jyu.fi}

\vspace{1cm}
{\em $^{\rm a}$ Department of Physics, University of Jyv\"askyl\"a,\\
P.O.Box 35, FIN-40351 Jyv\"askyl\"a, Finland\\}
\vspace{0.3cm}
{\em $^{\rm b}$ Department of Physics, University of Helsinki\\
P.O.Box 9, FIN-00014 University of Helsinki, Finland\\}
\vspace{0.3cm}
{\em $^{\rm c}$ Helsinki Institute of Physics,\\
P.O.Box 9, FIN-00014 University of Helsinki, Finland\\}

\vspace{1cm}
{\bf Abstract}

\end{centering}

\vspace{0.3cm}\noindent
We compute the centrality dependence of multiplicities of
particles produced in ultrarelativistic nuclear collisions
at various energies and atomic numbers. 
The computation is carried out in perturbative QCD with saturated 
densities of produced gluons and by including effects of nuclear geometry.
Numbers are given for Au+Au collisions at RHIC energies.

\vfill 

\end{titlepage}

\section{Introduction} 
The initial transverse energies and multiplicities in central
(zero impact parameter)
ultrarelativistic heavy ion collisions have been
computed in \cite{ekrt} from perturbative QCD supplemented
by the crucial assumption of 
saturation of produced semihard gluons:
a simple saturation criterion defines a saturation scale 
$\psat\sim 1$ (2) GeV at RHIC (LHC). Doing the computation at 
transverse momenta $p_T\ge\psat$
gives an estimate of the effect from all transverse momentum scales, 
both above and below $\psat$. ``Initial'' here then means proper
times of the order of $1/\psat\approx0.2$ fm (RHIC) and $\approx0.1$
fm (LHC). Assuming thermalisation at formation and 
further entropy conserving expansion, these initial
gluon numbers can be converted to final hadron multiplicities.
The predicted multiplicities agree well with the first
results from RHIC \cite{phobos}.

The results of \cite{ekrt} are formulated in the form of scaling rules,
quantity $\sim CA^a(\roots)^b$, in which the constants 
$C,a,b$ are determined for central $AA$ collisions.
A further variable one has experimental control over is the
centrality or impact parameter dependence of multiplicities and
transverse energies. In fact, in \cite{wg} the centrality
dependence of the multiplicity has been proposed as a means
of distinguishing between various models for particle production 
in ultrarelativistic heavy ion collisions.
In particular, in \cite{wg} one observed that 
the ratio of charged particle multiplicity to the number of
participants, $N_\rmi{ch}/N_{\rm part}$, grows when $b$
decreases while it decreases in the HIJING Monte Carlo
model \cite{hijing}.
The purpose of this note is a systematic study of the
behaviour of charged particle multiplicities with impact parameter
in the pQCD + saturation model.

\section{Impact parameter dependent saturation criterion}
Since the present work is a simple extension of \cite{ekrt}
to non-central $\bfb\not=0$ collisions, we refer there for a detailed
exposition of dynamical ideas.

The tools of describing non-central $AA$ collisions are \cite{ekl} the
nuclear density $n_A(r)$, normalised to $A$, the standard nuclear
thickness function $T_A(b)=\int dz n_A(r)$, normalised as $\int
d^2b\,T_A(b)=A$, and the nuclear overlap function $T_{AA}(b)$,
$\int\,d^2b\,T_{AA}(b)=A^2$, which is a 2-dimensional convolution in
$\bfb$-space of two $T_A(b)$'s. We shall systematically use the
physically relevant Woods-Saxon nuclear density distribution with
$R_A=1.12\,A^{1/3}-0.86\,A^{-1/3}$ fm, for which these integrals have
to be computed numerically.

Experimentally, however, one does not have a direct control over the impact
parameter. Instead, using forward calorimeters one can measure the
number of participants $N_\rmi{part}$ in the collision; 
for central collisions $N_\rmi{part}\lsim2A$.  
This can be approximately converted to
impact parameter by the formula 
\ba
N_\rmi{part}(b) = \int\,d^2s\,T_A(\bfb-\bfs)
\left[1-\exp(-\sigma_\rmi{in}T_B(\bfs))\right] + 
\int\,d^2s\,T_B(\bfs)
\left[1-\exp(-\sigma_\rmi{in}T_A(\bfb-\bfs))\right] 
\nonumber
\ea
\be
\hspace{-3cm}=2\int\,d^2s\,T_A(\bfb-\bfs)  
\left[1-\exp(-\sigma_\rmi{in}T_A(\bfs))\right] \quad (A=B),  
\la{npartvsb} 
\ee
where $\sigma_\rmi{in}\approx 40$ mb is the inelastic pp cross
section in the RHIC energy range.
An example is shown in Fig.~\ref{nvsb}. The Poissonian
$\exp(-\sigma_\rmi{in} T_A)$ in \nr{npartvsb} can as well be replaced
by a binomial-like $(1-\sigma T/A)^A$; the numerical effect is
negligible. 
We have taken $\sigma_\rmi{in}= 35$, 39 and 42 mb for $\sqrt s=56$, 130
and 200 GeV, respectively, based on \cite{pdg}.

We have now the tools to generalise the $\bfb=0$ saturation condition
for initially produced partons,
\be 
N_{AA}(p_0,\bfb=0)=2T_{AA}(0)\sigma_\rmi{pQCD}(p_0)=
\sum_{k=g,q,\bar q}N_k(p_0)=p_0^2R_A^2, 
\la{psat} 
\ee
to arbitrary $\bfb$. Perturbative
QCD enters via an LO computation of the inclusive production cross
section $\sigma_\rmi{pQCD}(p_0)$ of (mini)jets with $p_T\ge p_0$
and $|y|\le0.5$, with NLO
contributions taken into account via an overall $K$-factor $K=2$ according 
to \cite{et}; all the formulas are spelled out in detail in \cite{ek96}.
Eq. \nr{psat} expresses the fact that at saturation $N_{AA}(p_0=\psat)$
quanta each with transverse area $\pi/\psat^2$ fill the whole nuclear
transverse area $\pi R_A^2$. Numerical, group theory factors or powers
of $g$ could be included \cite{mueller99}, but these are anyway ${\cal
O}(1)$ unless one discusses a parametric weak coupling limit
$g\to0$. All parton flavours are included, though at $p_0=\psat$
gluons clearly dominate even at lowest energies.

\begin{figure}[htb]
\hspace{0.5cm}
\epsfysize=12cm
\centerline{\epsffile{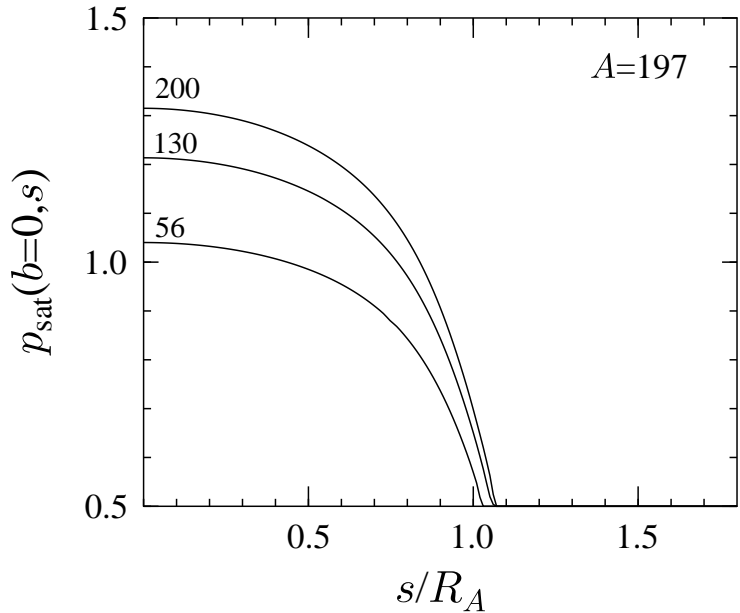}}
\vspace{-2cm}
\caption[a]{\protect \small The saturation momentum $\psat(b=0,s)$ computed
as a solution of Eq.\nr{psatb} for $\bfb=0$,
$A=197$ and $\roots=56,130,200$ GeV.
}
\la{psatb0}
\end{figure}

In terms of an average transverse area density in central collisions,
$dN_{AA}/d^2s \approx N_{AA}(p_0,{\bf 0})/(\pi R_A^2)$, Eq. \nr{psat} becomes
$dN_{AA}/d^2s \cdot \pi/p_0^2=1$. For arbitrary $\bfb\,$  and \bfs, the average
saturation criterion thus generalizes to a local one,
\be
{dN_{AA}(p_0,\bfb,\bfs)\over d^2s}=
2 T_{A}(\bfb-\bfs)T_A(\bfs)\sigma_\rmi{pQCD}(p_0)=
p_0^2/\pi.
\la{psatb}
\ee
We shall choose the coordinate system at $\bfb\not=0$ so that
$\bfb=(b,0)$ and so that 
the centers of the nuclei are at $(b/2,0)$ and $(-b/2,0)$. 
A solution of Eq.\nr{psatb} gives us the local saturation momentum
$\psat(b,\bfs)$ at the position $\bfs$ of an A+A collision at
impact parameter $\bfb$. Fig.~\ref{psatb0} shows $\psat(b=0,\bfs)$,
which only depends on $s$;
Fig.~\ref{psatb1} shows $\psat(b=R_A,\bfs=(s_x,0))$
and  $\psat(b=R_A,\bfs=(0,s_y))$.

\begin{figure}[htb]
\vspace{-1.5cm}
\hspace{0.5cm}
\epsfysize=12cm
\centerline{\epsffile{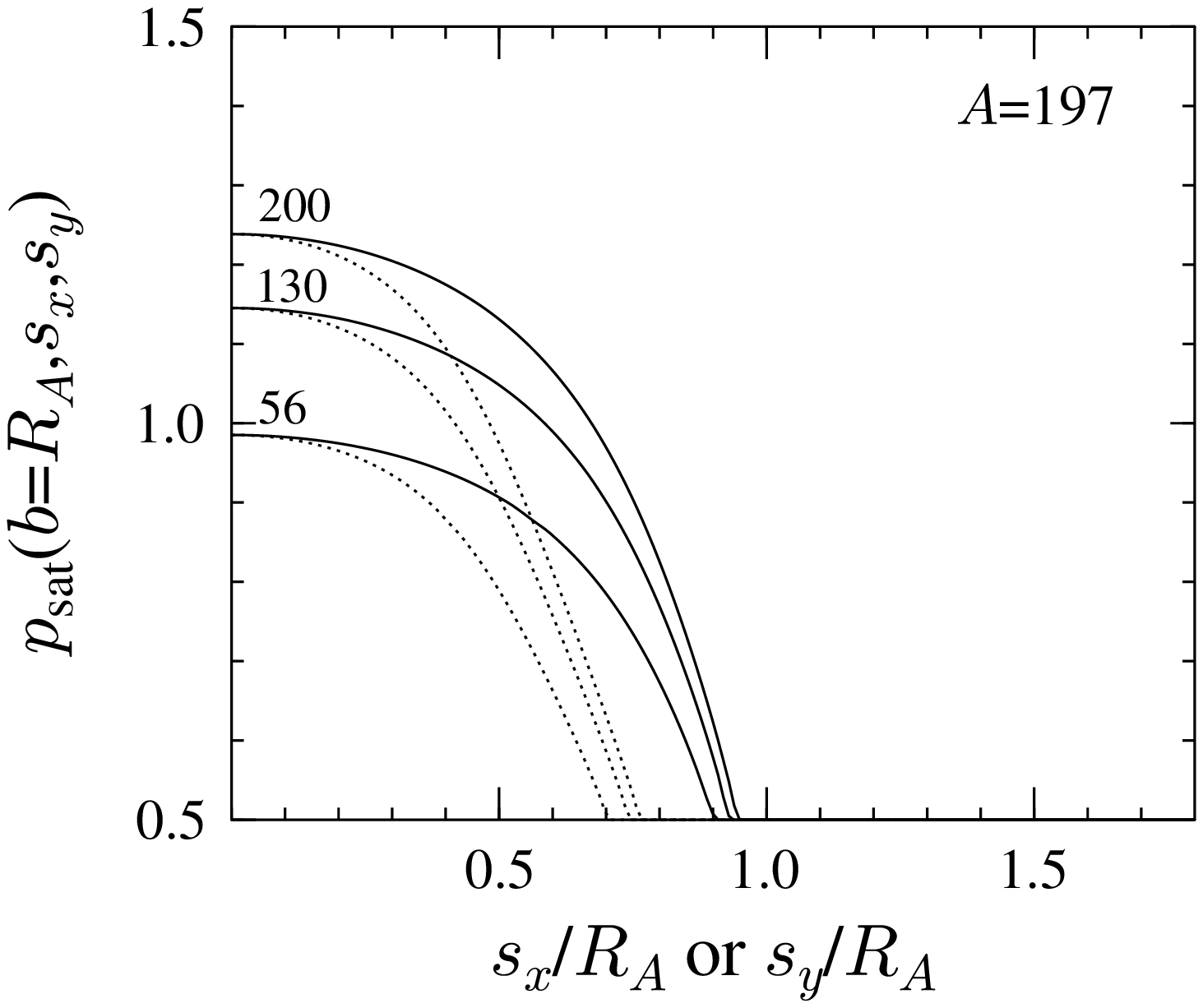}}
\vspace{-2.5cm}
\caption[a]{\protect \small The saturation momentum 
$\psat(b=R_A,\bfs)$ along the symmetry axes of the overlap
area. Solid curves are for $\bfs=(s_x,0)$ and the dotted ones for 
$\bfs=(0,s_y)$,
computed from Eq. \nr{psatb} for a Woods-Saxon
nuclear density distribution for $A=197$ and 
$\roots=56,130,200$ GeV.}  
\la{psatb1}
\end{figure}

The predicted initial multiplicity at fixed $b$ then simply is
$dN_{AA}/d^2s$ at the solution of \nr{psatb} integrated over $\bfs$.
In terms of the right-hand side: \be N_{AA}(b)=\int
d^2s\,\psat^2(b,\bfs)/\pi.  \la{naab} \ee Here an important issue
enters: how does one deal with large values of $s$ or small values of
$\psat(b,\bfs)$ in \nr{naab}? Physically, how does one treat very
peripheral collisions for which $\psat$ becomes small,
nonperturbative?  One may note that even at SPS energies, where
$\psat\approx0.7$ GeV, the data was reproduced by this model up to a
20\% error.  Clearly the pQCD+saturation model can be reliable only if
the large $s$, small $\psat$ region makes a negligible contribution to
the integral \nr{naab}. This, in fact, is the case: in central
collisions, for example, the contribution of the range from
$\psat\ge0.7$ GeV to $\psat\ge0.5$ GeV to the integration \nr{naab} is
only an $\approx$ 4\% increase in $N_{AA}$.  In Figs. \ref{psatb0} and
\ref{psatb1} the decrease in $\psat^2$ for $\psat<0.5$ GeV is very
rapid and the surface integral \nr{naab} is negligibly affected.
Values of $N_{AA}(b)$ as computed from \nr{naab} including only
$\psat\ge0.5$ GeV are plotted in Fig.~\ref{nvsb}.  The total initial
particle production becomes thus described by the saturation model in
an effective way in the sense that by pushing the local saturation
criterion to its limits (to $\psat\rightarrow 0$) no additional,
non-saturated, components need to be included.  For very peripheral
collisions, or for pp collisions, the model cannot be applied.

The values of $b$ in Figs. \ref{psatb0} to \ref{nvsb} can be converted
to the experimentally more directly accessible number $N_\rmi{part}$
using Eq. \nr{npartvsb} and Fig.~\ref{nvsb}. Concretely,
Figs. \ref{psatb0} to \ref{nvsb} extend to $b=1.8R_A$, which
corresponds to $N_\rmi{part}\approx30$; similarly $b=1.6R_A$ would
correspond to $N_\rmi{part}\approx50$. For still larger $b$ one
expects saturation effects to disappear
and this region is thus beyond the scope of the current study.

\begin{figure}[htb]
\vspace{-2cm}
\hspace{0.5cm}
\epsfysize=12cm
\centerline{\epsffile{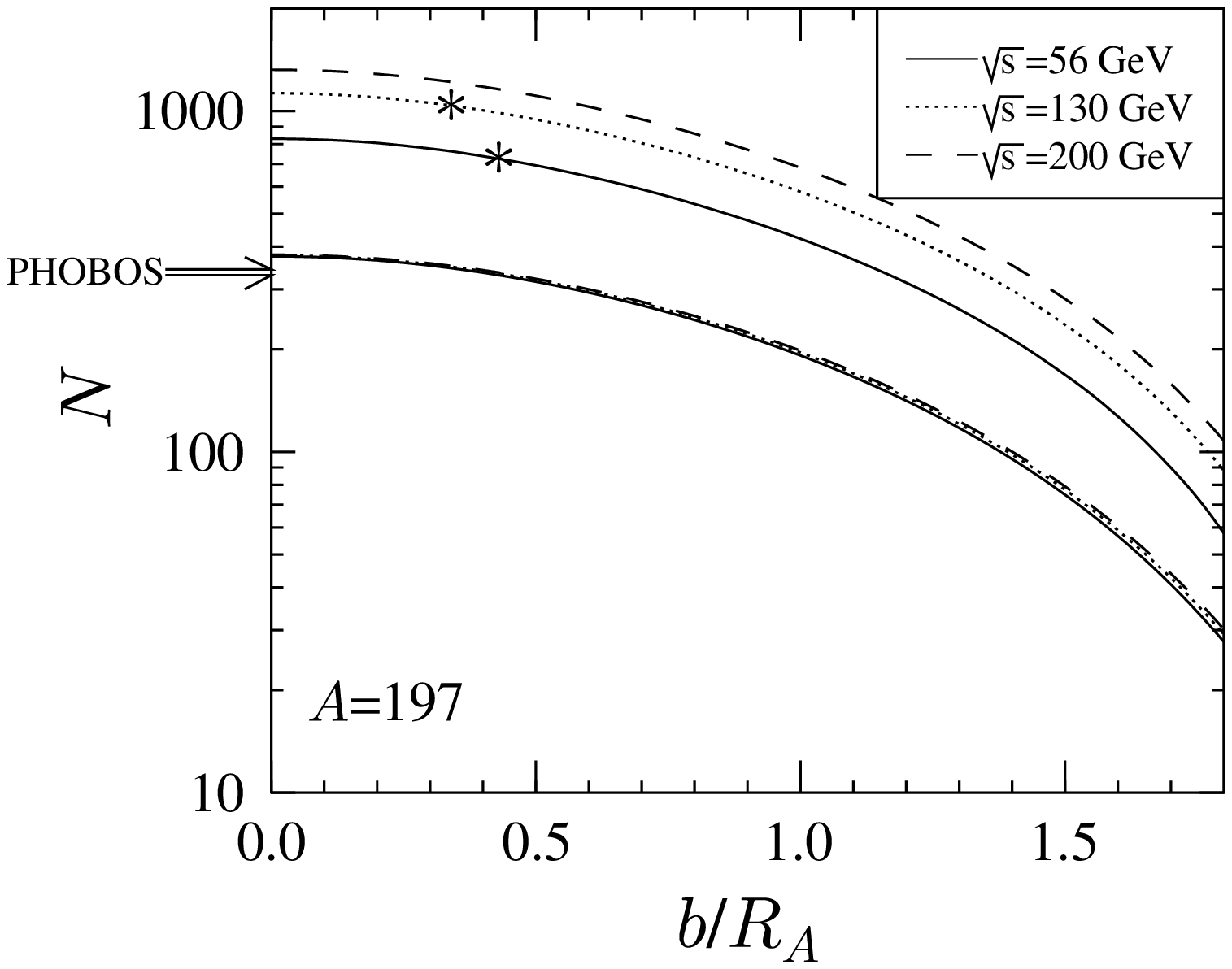}}
\vspace{-2.5cm}
\caption[a]{\protect \small The initial ($\tau\approx1/\psat$) 
number of quanta
$N_{AA}(b)$ produced in $^{197}$Au+$^{197}$Au collisions
at $\roots=$56, 130 and 200 GeV as a function of $b/R_A$ (the upper set of 
three curves), as computed from Eq.\nr{naab} with $\psat\ge0.5$ GeV.
The three lower curves show $N_\rmi{part}(b)$ computed from 
\nr{npartvsb} for $\sigma_\rmi{in}= 35,39,42$ mb (bottom to top). The
average values of $N_\rmi{part}(b)$ given by PHOBOS 
\cite{phobos} are indicated 
by the two (almost overlapping) arrows on the vertical axis, 
the values of $N_{AA}(b)$ at the
corresponding values of $b/R_A$ are marked by the asterisks. 
}
\la{nvsb}
\end{figure}

We shall here not discuss the transverse energy $E_T$ production in
detail, since for it the relation between the initially produced
and finally observed $E_T$ per unit rapidity is likely to be more
complicated than that for multiplicity, due to 
$pdV$ work and transverse flow effects in the course of
expansion. However, a reasonable estimate for the
initial production is the same $E_T$ per quantum as at $\bfb=0$:
$E_T(b)/N(b) \approx 1.35\,\langle\psat(b)\rangle$, where
$\langle\psat(b)\rangle$ is computed with a weight 
$T_A(\bfs)T_{A}(\bfb-\bfs)$ for  $\psat(b,\bfs)$.
Note also that it is here that these
pQCD+saturation results deviate significantly from the
classical field computations \cite{kv1,kv2}: their $E_T/N$ is
larger by a factor 3.

\section{Comparison with experiment}
The above were predictions for the multiplicity
integrated along a curve $\tau_i\approx1/\psat(b,\bfs)$
of initial time. Comparing the multiplicities with experiment requires
assumptions on the expansion and decoupling of the system. These
questions are discussed extensively in the literature and one
is looking forward to experimental tests. 

We shall here use the simplest assumption of 
early thermalisation and entropy conserving longitudinally 
boost invariant expansion. 
At the central slice $z=0$ the local particle density is initially
then $n(\tau_i(b,\bfs),\bfs)\approx dN/d^2s/(\tau_i\Delta
\eta)={\cal S}(\tau_i,s)/3.6$, where $\eta$ is the space-time
rapidity and ${\cal S}$ is the entropy density. The initial total
entropy per rapidity unit, $S_i$, can thus be computed from $S_i(b)=3.6
N_{AA}(b)$.
This  predicts that 
the relation of the number of initially produced particles (gluons)
to the total number of particles in the final state (pions per unit rapidity)
is  $N_f/N_{AA}=(N_f/S_f)/(N_{AA}/S_i)*(S_f/S_i)\approx3.6/4$, inserting
the number/entropy ratios for massless gluons and massive pions. 
To obtain the experimentally measured $dN_\rmi{ch}/d\eta$, 
$\eta$=pseudorapidity, we use the estimates
$N_\rmi{ch}\approx \fr23 N_f$ and $dN/d\eta\approx0.9 dN/dy$.
Fig.~\ref{nvspartpairs} then shows the quantity 
$(dN_{\rm ch}(b)/d\eta)/(N_\rmi{part}/2)$, i.e., the number of 
charged particles produced in unit pseudorapidity
per number of participant pairs, as a function 
of the number of participants. The parameters
($A$ and $\sqrt s$) are those for RHIC experiments.

\begin{figure}[htb]
\vspace{-2cm}
\hspace{0.5cm}
\epsfysize=12cm
\centerline{\epsffile{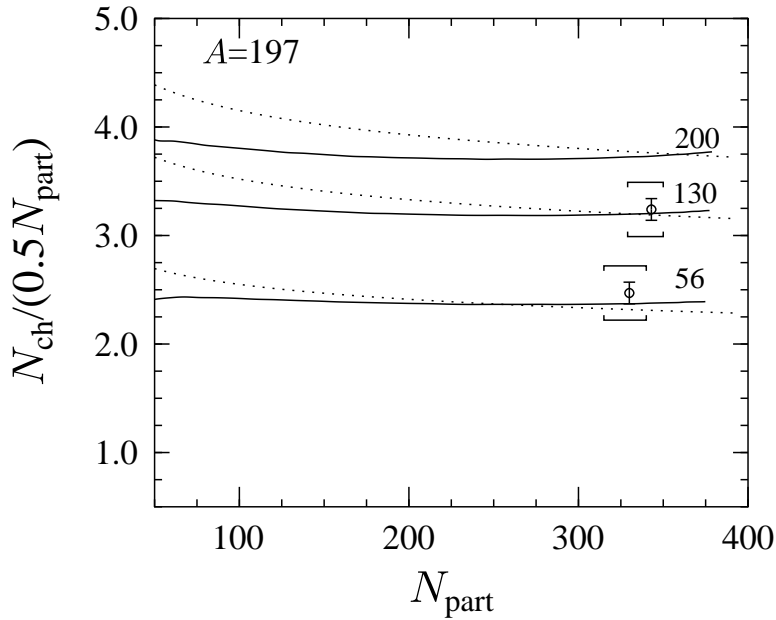}}
\vspace{-2.5cm}
\caption[a]{\protect \small Solid curves: the final number of charged
particles per unit pseudorapidity and per participant pair 
produced in Au+Au
collisions at $\roots=56$, 130, 200 GeV vs the number of
participants.  Dotted curves: the same for central collisions with
effective nuclei $A_{\rm eff}=N_{\rm part}/2$. The PHOBOS data for 56
and 130 GeV are also shown, with separate error bars for statistical
and systematic errors \cite{phobos}.  
}  
\la{nvspartpairs}
\end{figure}

For central collisions (approximately; as shown in Fig.\ref{nvsb},
the number of participants in
\cite{phobos} is somewhat less than 2A) the agreement with
experiment is good. The rest of the curve is a prediction. 

The prediction in Fig.\ref{nvspartpairs} has the
striking feature of being essentially constant in $N_\rmi{part}$.
This is in agreement with the estimate in
\cite{wg}, shown by the dotted lines. 
The small difference is due to the fact that in \cite{wg}
one did not actually
perform the saturation computation at arbitrary $\bfb$ but instead
used $\bfb=0$ results at $A_\rmi{eff}=N_\rmi{part}(\bfb)/2$. 
Due to the computed \cite{ekrt} $A$-scaling of $N_\rmi{AA}$ at $\bfb=0$,
$N_\rmi{AA}\sim A^{0.922}$, this immediately results in
$N_\rmi{AA}/N_\rmi{part}\sim 1/N_\rmi{part}^{0.08}$, 
a slow increase as $N_\rmi{part}$ decreases. At very small $N_\rmi{part}$
this would seem to lead to a striking effect, but then one enters
the very peripheral region in which saturation does not work.

\section{Conclusions}

In this note we have extended the pQCD + saturation model for initial
production of partons in $A+A$ collisions, studied in quantitative
detail in \cite{ekrt}, from central ($\bfb=0$) to non-central
collisions, up to perhaps $b=1.6R_A$.
This extension is based on a simple geometric
argument. Subject to the usual uncertainties in relating
initial to final observed numbers, predictions for
particle multiplicities and transverse energies can now be made
and tested in terms of three variables, $A$, $\roots$ and $b$ 
(or number of participants).

It is impossible to give a systematic estimate of the theoretical
error in the predictions. Corrections to pQCD can, in principle, be
computed, but the use of the saturation scale adds a nonperturbative,
even phenomenological, element. What one can say is that the model is
unambiguously determined, fits the data at SPS within 20\% and predicted
well the first measurements at RHIC, so maybe it is an acceptable way
of at least extrapolating to so far unobserved parameter values.
It thus offers a strongly constrained framework for an 
analysis of experimental data in
a 3-parameter ($A$,$\roots$, $b$) space and one should be able
to make a distinction between different models along the lines
suggested in \cite{wg}.

{\bf Acknowledgements} We thank M. Gyulassy, V. Ruuskanen and X.-N. Wang 
for discussions and advice. Financial support from the Academy of Finland and 
Waldemar von Frenckell Foundation is gratefully acknowledged.


\begin{thebibliography}{9}

\bibitem{ekrt}
K.~J.~Eskola, K.~Kajantie, P.~V.~Ruuskanen and K.~Tuominen, 
Nucl.\ Phys.\  {\bf B570} (2000) 379
[hep-ph/9909456].


\bibitem{phobos}
B.~B.~Back {\it et al.}  [PHOBOS Collaboration],
``Charged particle multiplicity near mid-rapidity in 
central Au + Au  collisions at $\sqrt s$ = 56  and 130 AGeV,''
hep-ex/0007036.

\bibitem{wg}
X.-N.~Wang and M.~Gyulassy,
``Energy and centrality dependence of rapidity densities at RHIC,''
nucl-th/0008014.

\bibitem{hijing}
X.-N.~Wang and M.~Gyulassy,
Phys.\ Rev.\  {\bf D44} (1991) 3501.

\bibitem{ekl}
K.~J.~Eskola, K.~Kajantie and J.~Lindfors,
Nucl.\ Phys.\  {\bf B323} (1989) 37.

\bibitem{pdg} The Review of Particle Physics, The European 
Physical Journal {\bf C15} (2000) 1, Fig. 37.19.

\bibitem{et}
K.~J.~Eskola and K.~Tuominen,
``Production of transverse energy from minijets in next-to-leading 
order  perturbative QCD,''
hep-ph/0002008, Phys. Lett. {\bf B}, in press.

\bibitem{ek96}
K.~J.~Eskola and K.~Kajantie,
Z.\ Phys.\  {\bf C75} (1997) 515
[nucl-th/9610015].

\bibitem{mueller99}
A.~H.~Mueller,
Nucl.\ Phys.\  {\bf B572} (2000) 227
[hep-ph/9906322].

\bibitem{kv1}
A.~Krasnitz and R.~Venugopalan,
``The initial gluon multiplicity in heavy ion collisions,''
hep-ph/0007108.

\bibitem{kv2}
A.~Krasnitz and R.~Venugopalan,
Phys.\ Rev.\ Lett.\  {\bf 84} (2000) 4309
[hep-ph/9909203].

\end{thebibliography}
\end{document}